\documentclass[aps, prb,10pt,preprint,amsmath,amssymb, 
                 superscriptaddress,longbibliography]{revtex4-1}

\usepackage{lineno}
\usepackage[]{graphicx}
\usepackage{wrapfig}
\usepackage[]{verbatim}
\usepackage[dvipsnames]{xcolor}

\usepackage{bm}
\usepackage{graphicx}
\usepackage[breaklinks,colorlinks=true,citecolor=blue]{hyperref}
\usepackage{mathrsfs}
\usepackage[lofdepth,lotdepth,caption=false]{subfig}
\usepackage{varwidth}
\usepackage{wrapfig}
\usepackage{times}
\usepackage{longtable}
\usepackage{multirow}
\usepackage{tikz}

\begin{document}

\title{Iterative peak-fitting of frequency-domain data via deep convolution neural networks}

\author{Seong-Heum Park}
\affiliation{Department of Physics, Kangwon National University, Chuncheon 24341, Korea}

\author{Hyeongseon Park}
\affiliation{Department of Physics, Kangwon National University, Chuncheon 24341, Korea}

\author{Hyunbok Lee}
\email{hyunbok@kangwon.ac.kr}
\affiliation{Department of Physics, Kangwon National University, Chuncheon 24341, Korea}
\affiliation{Institute for Accelerator Science, Kangwon National University, Chuncheon 24341, Korea}

\author{Heung-Sik Kim}
\email{heungsikim@kangwon.ac.kr}
\affiliation{Department of Physics, Kangwon National University, Chuncheon 24341, Korea}
\affiliation{Institute for Accelerator Science, Kangwon National University, Chuncheon 24341, Korea}

\begin{abstract}
High-throughput material screening for the discovery and design of novel functional materials requires automatized analyses of theoretical and experimental data. Here we study the subject of human-free analyses of one-dimensional spectroscopic data, {\it e.g.} in the frequency domain, via employing deep convolution neural network. Specifically, we trained various deep convolution neural network and benchmarked their performance in decomposing one-dimensional noisy data into multiple nonorthogonal peaks in an iterative manner, after which a conventional basin-hopping algorithm was applied to further reduce residual fitting error. Among six different network architectures, a variant of ``Squeeze-and-excitation" network (SENet) structure that we first propose in this study shows the best performance. Dependency of training performance with respect to the choice of the loss function is also discussed. We conclude by applying our modified SENet model to experimental photoemission spectra of graphene, MoS$_2$, and WS$_2$ and address its potential applications and limitations. 
\end{abstract}


\maketitle

\section{Introduction}

Thanks to recent advancements in the capabilities of synthesizing, characterizing, and analyzing new materials, there has been a surge of theoretical and experimental data on potentially interesting systems that has accumulated over the last decade. Specifically, developments in {\it ab-initio} electronic structure calculation methods such as density functional theory\cite{DFT_Review1,DFT_Review2} have enabled the compilation of extensive computational materials databases such as {\sc materials project}\footnote{https://materialsproject.org}, {\sc open quantum materials database (OQMD)}\footnote{http://oqmd.org}, or {\sc novel materials discovery (NOMAD)}\footnote{https://nomad-coe.eu}, based on which data mining and machine-learning approaches can be applied to establish statistical models with enough predicting power for novel materials discovery and inverse materials design\cite{datascience2017,CGCNN2018,Butler2018,Zunger2018,Sanchez-Lengeling360,Schleder_2019,Himanen_2019}. 

A relatively under-explored area, compared to analyzing and learning from cumulated computational data for training predictive models, is applying machine learning for analyses of experimental spectroscopic data. There have been a few earlier works on the application of machine learning in distilling useful physical information from experimental photoemission and scanning tunneling microscopy spectra\cite{Zhang2019,Drera_2020,Youhei_2019}, and specifically a recent study on application of deep neural network to core-level x-ray photoemission spectra\cite{Drera_2020} showed a qualitative agreement with experimental results in quantifying stoichiometry of chemical elements of given test compounds, but with about 10\% error in actual composition ratio. Compared to its importance in condensed matter and materials sciences studies, however, the number of subsequent studies has been rather limited. Several difficulties in analyzing experimental spectroscopic data are {\it i)} inherent non-uniqueness in decomposing spectra into non-orthogonal elements (such as peaks), {\it ii)} limited amount of experimental spectra which are often insufficient to train deep neural networks, and {\it iii)} superiority of human recognition based on physical intuitions over machine-trained models in traditional spectroscopy studies. 

Since there are a number of spectroscopic tools where the data is measured in energy or frequency domain, such as, not limited to, Raman, photoemission, light absorption, photoluminescence, nuclear magnetic resonance, and inelastic x-ray or neutron scatterings, computational tools to automatize analyses of experimental spectra are essential in order to enable data-driven materials scientific studies. For this purpose, here we combined deep convolution neural networks (CNN) and conventional basin-hopping optimization techniques to decompose a given spectrum $F(\omega)$ in one-dimensional parameter ($\omega$) domain into multiple peak functions $f_i(\omega)$ ($i$ being a peak label), so that the error between original $F(\omega)$ and the sum of $f_i(\omega)$ becomes minimized with minimal human intervention. Specifically, in search of the best CNN model to tackle the non-unique fitting problem, we employed six deep CNN architectures with varying network topology and complexity; 5 of them have been recently suggested, while the last one (modified SENet, denoted as m-SENet hereafter) is newly proposed in this work. We found that, among the six proposed CNN architectures employed, our m-SENet shows the best fitting performance. 

\begin{figure}
  \centering
  \includegraphics[width=0.7\textwidth]{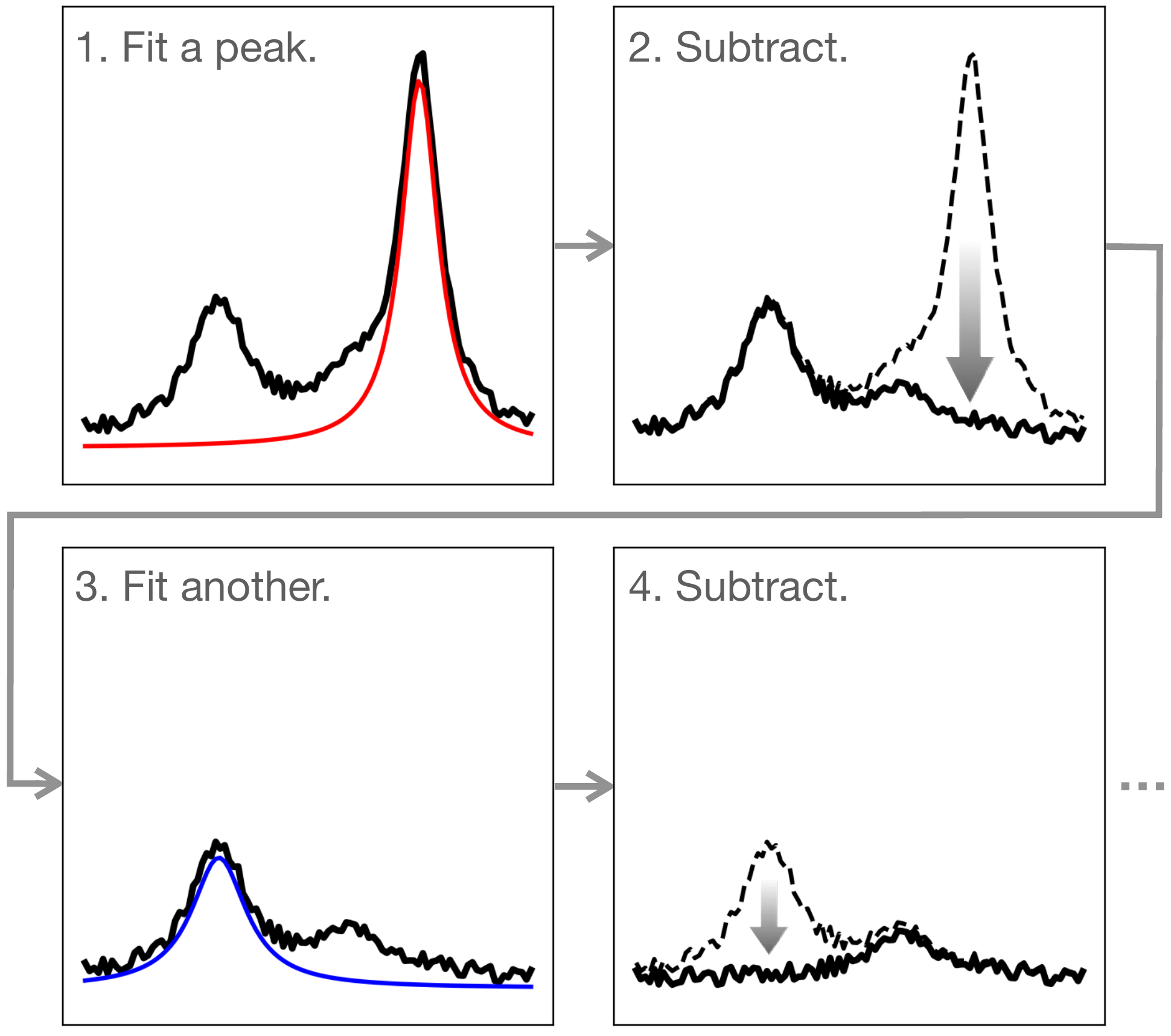}
  \caption{
  A cartoon illustration of decomposing a noisy spectrum (black curve) with multiple peak features by sequentially applying regression and subtraction of the largest-area peak (red and blue curves) until no peak feature is left. 
  }
  \label{fig:fitting}
\end{figure}

Our trained CNN models were then used to find a peak with the largest underlying area so that the fitted peak can be subtracted repeatedly until no peak feature is left in the spectrum; see an illustration of how this works in Fig.~\ref{fig:fitting}, where the trained CNN almost exactly captures peak features. Limitation in the size of accessible experimental spectroscopy data was circumvented by using synthetic spectra generated via the pseudo-Voigt function\cite{Ida:nt0146}. Finally, errors that occur in the regressions process are corrected after all peaks are extracted by applying conventional global optimizations such as basin-hopping algorithm, which fine-tunes CNN-extracted peak parameters and minimizes the loss function. After checking the fitting performance of our model, we applied our best trained model to actual experimental photoemission data from graphene, MoS$_2$, and WS$_2$ and confirmed the predicting power of our approach in fitting the actual experimental spectroscopy data. We conclude by discussing potential loopholes of our approach and necessary improvements.

\section{Methods}

\subsection{Description of neural network architectures}
As mentioned in the Introduction section, we need a well-trained CNN architecture to accurately subtract a peak with the largest underlying area from a spectrum. For this purpose, we benchmarked fitting performances of six convolution neural network (CNN) architectures; i) LeNet \cite{lecun1998gradient,lecun1989backpropagation}, Alex-ZFNet \cite{krizhevsky2012imagenet,zeiler2014visualizing}, VGGNet \cite{simonyan2014very}, ResNet \cite{he2016deep}, SENet \cite{hu2018squeeze}, and a modified version of SENet (hereafter denoted as m-SENet), listed in order of increasing complexity in network structure. LeNet is the first proposed CNN architecture, which consists of alternating convolution and subsampling (pooling) layers. Alex-ZFNet\cite{krizhevsky2012imagenet,zeiler2014visualizing} can be considered an improvement of LeNet by introducing maxpooling, dropout, and rectified linear unit (ReLU) activation function. VGGNet\cite{simonyan2014very} takes a different path by adopting a uniform network structure but with increased number of convolution filters. ResNet\cite{he2016deep} employs the so-called ``skip connections" to enhance back-propagation of the loss function gradient deep into the network layers. In SENet\cite{hu2018squeeze} a so-called 'squeeze-and-excitation (SE) block' is introduced to improve performance with a small amount of computational cost increase. 

\begin{figure}
  \centering
  \includegraphics[width=0.95\textwidth]{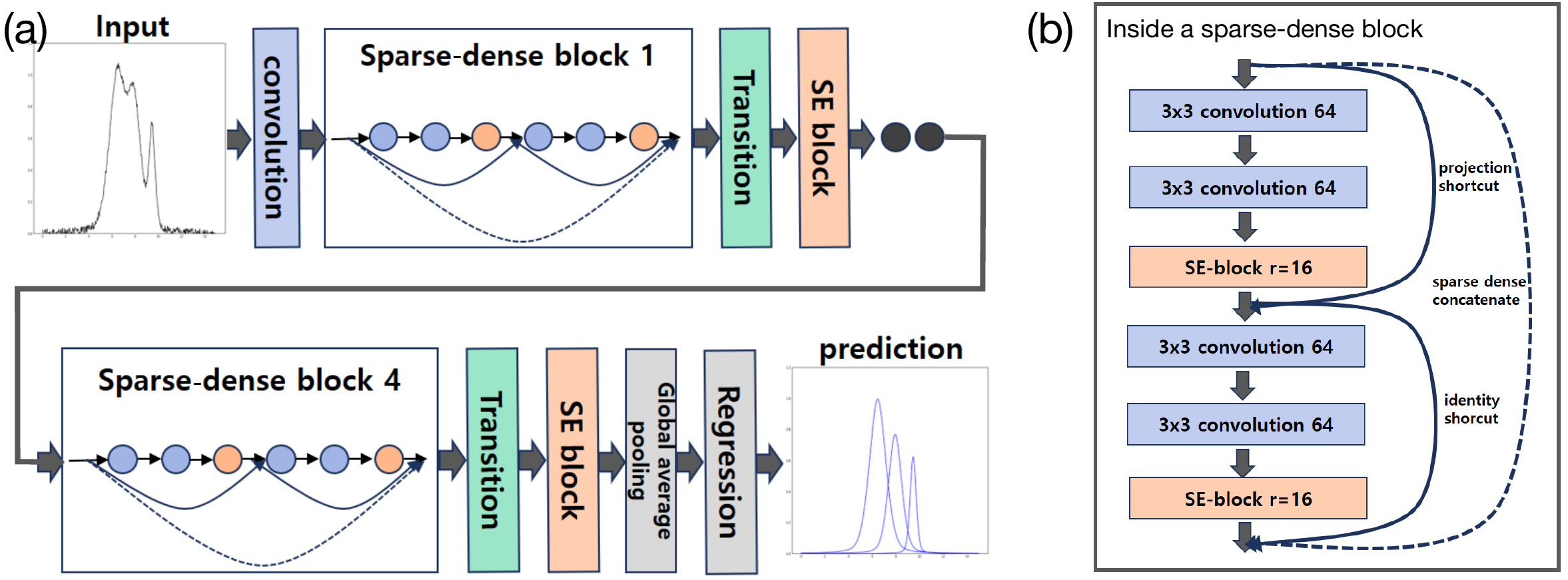}
  \caption{
  (a) A schematic representation of m-SENet architecture, where detailed layering structure inside a sparse-dense block is depicted in (b). 
  }
  \label{fig:struct}
\end{figure}

For this work we employed the m-SENet (see Fig.~\ref{fig:struct}), where the idea of SE block from SENet was implemented into the so-called 'sparse-dense block' as shown in Fig.~\ref{fig:struct}(a) and (b). Each sparse-dense block consists of serially connected pairs of convolution layers (CONV) and a SE block, where identity mapping connects input and output layers of two CONV-CONV-SE triplets in a sparse-dense block. Batch normalization and SE layers were placed between adjacent sparse-dense blocks to rebalance weights of convolution feature maps. This structure allows efficient learning even in extremely deep models. Finally, the regression layer was attached through global average pooling. The number of total trainable parameters was 4,183,830 and ‘overlapped average pooling’ was used in all subsampling processes.

\subsection{Generation of training, validation, and test datasets}
Compared to the large amount of dataset required to train deep neural networks, the amount of accumulated spectroscopic data from experimental measurements is often insufficient. Hence we generated a synthetic dataset for training networks employing pseudo-Voigt profile\cite{Ida:nt0146}, which is often adopted for fitting peaks from experimentally measured x-ray diffraction or photoemission data. The following form was chosen to generate each peak in the artificial dataset, 
\begin{equation}
f(\omega ; \omega_0, I_0, \delta) = 
I_0 \left( 
I_G e^{-\frac{\log(2) (\omega - \omega^2)}{\delta^2 \omega^2_G}} + I_L \frac{1}{1+\frac{(\omega-\omega_0)^2}{\delta^2 \omega^2_L}}
\right),
\label{eq:psVoigt}
\end{equation}
where $\omega$ can be considered frequency (energy) in spectroscopic data, with $\omega_0$, $I_0$, and $\delta$ being randomly generated position, maximum height, and (dimensionless) width of the peak, respectively.\footnote{$\omega_0$, $I_0$, and $\delta$ are randomly chosen in the range of $0.13 \omega_{\rm max} < \omega_0 < 0.87 \omega_{\rm max}$,  $0.02 \omega_{\rm max} < \delta < 0.13 \omega_{\rm max}$, and $0.05 < I_0 < 1.05$, respectively, where $0 \leq \omega \leq \omega_{\rm max}$ is our frequency domain.} $\omega_G$, $\omega_L$, $I_G$, and $I_L$ are set to be 0.510, 0.441, 0.7, and 0.3, respectively; this choice of parameters in the above pseudo-Voigt profile was adopted to fit experimental spectra in a previous photoemission study\cite{Kim_2018}.

\begin{figure}
  \centering
  \includegraphics[width=0.80\textwidth]{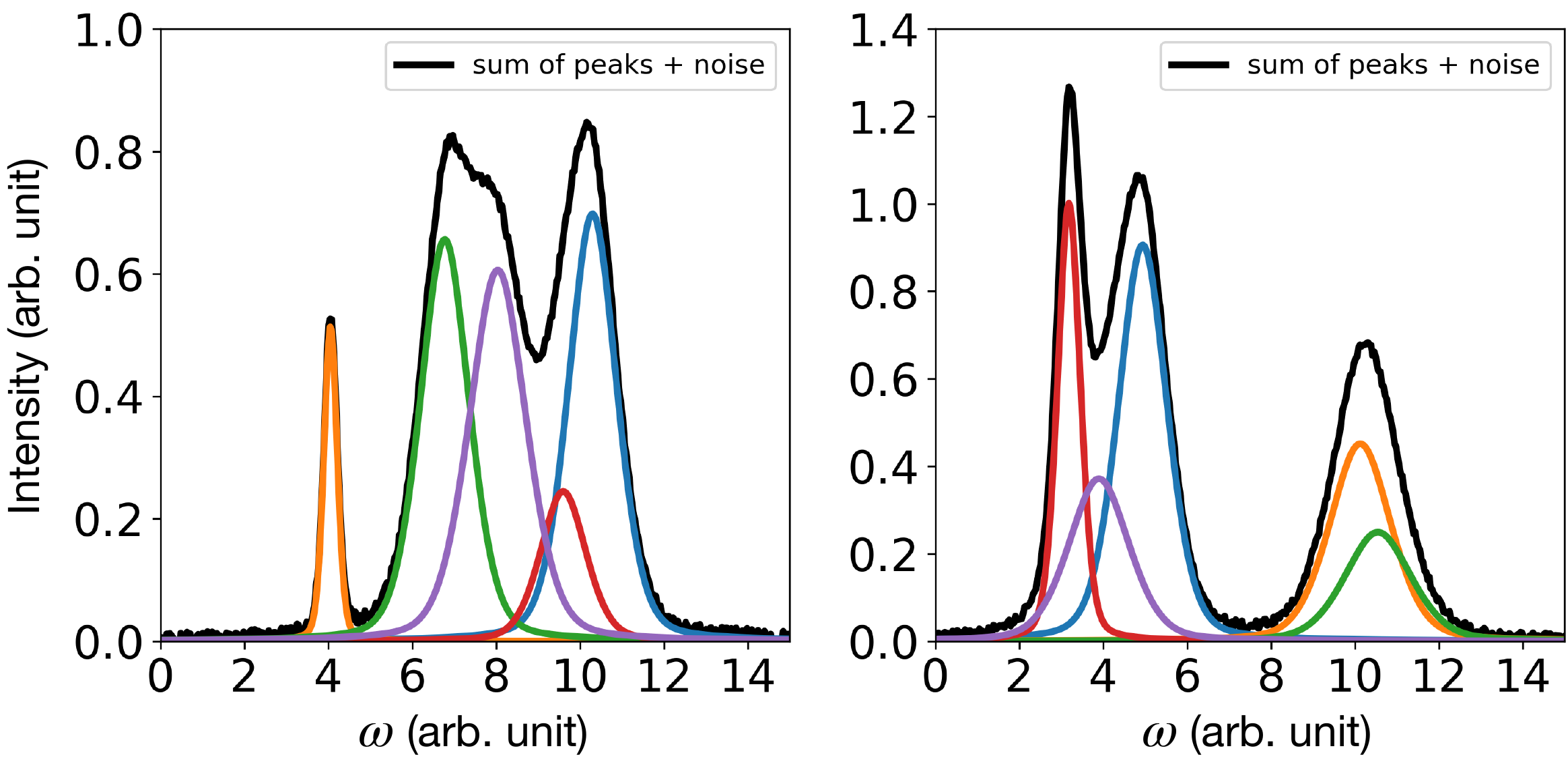}
  \caption{
  Two examples of synthetically generated data with multiple (up to 5) pseudo-Voigt profiles. Colored and black lines depict each pseudo-Voigt peaks and addition of all peaks with gaussian noise, respectively. 
  }
  \label{fig:dataexample}
\end{figure}

In each one-dimensional synthetic data up to 5 random pseudo-Voigt functions were added, with the addition of gaussian noise of which standard deviation being 2\% of the maximum allowed $I_0$ (see Fig.~\ref{fig:dataexample}). We also enforced centers of neighboring peaks to be no closer than 20\% of the maximum allowed $\delta$. Background signals other than gaussian noises are not considered in this work, however, it was previously reported that filtering monotonically increasing or decreasing background signals can be easily done employing a simple CNN network.\cite{PeakFitting}

\subsection{Implementation, train, and validation of neural networks}
To implement various CNN structures and train them we employed {\sc keras}\cite{chollet2015keras} framework and {\sc tensorflow}\cite{tensorflow2015-whitepaper} backend. Training of CNNs was done using a Nvidia RTX 2080Ti graphic processing unit. Four different target variables --- center $\omega_0$, width $\delta$, and amplitude $I_0$ of the peak with the largest underlying area, and the number of peaks in the spectrum --- contribute to the total loss function, which is the sum of loss functions for center, width, amplitude, and peak numbers. To regularize contributions of different loss components in the total loss, weights of 1, 10, 20, and 2 were assigned to each part of the loss function optimizing $\omega_0$, $\delta$, $I_0$, and peak number, respectively. {\sc adam}\cite{kingma2014adam} optimizer was employed throughout the whole training and test jobs for this study. The stochastic gradient descent (SGD) method\cite{SGD-Bottou} was tested in comparison to {\sc adam} results and found to show almost the same optimization performance compared to {\sc adam} in this study. In all cases, LeakyReLU activation function was used\cite{xu2015empirical}, and trainings were done during 50 epochs with 512 batch-size. The learning rate was reduced to 10\% after 20th and 40th epochs to avoid overfitting. No dropout layer was used in this study. 

For the CNN benchmark results presented in this work, we generated 1.5$\times$10$^6$ synthetic spectra, among which 1.2$\times$10$^6$, 1.5$\times$10$^5$, 1.5$\times$10$^5$ data were used for training model, validating and testing model performance, respectively. Later we increased the size of the dataset to 10$^7$ to check performance improvements.

\section{Results}

\subsection{CNN training and validation}

\begin{figure}
  \centering
  \includegraphics[width=0.70\textwidth]{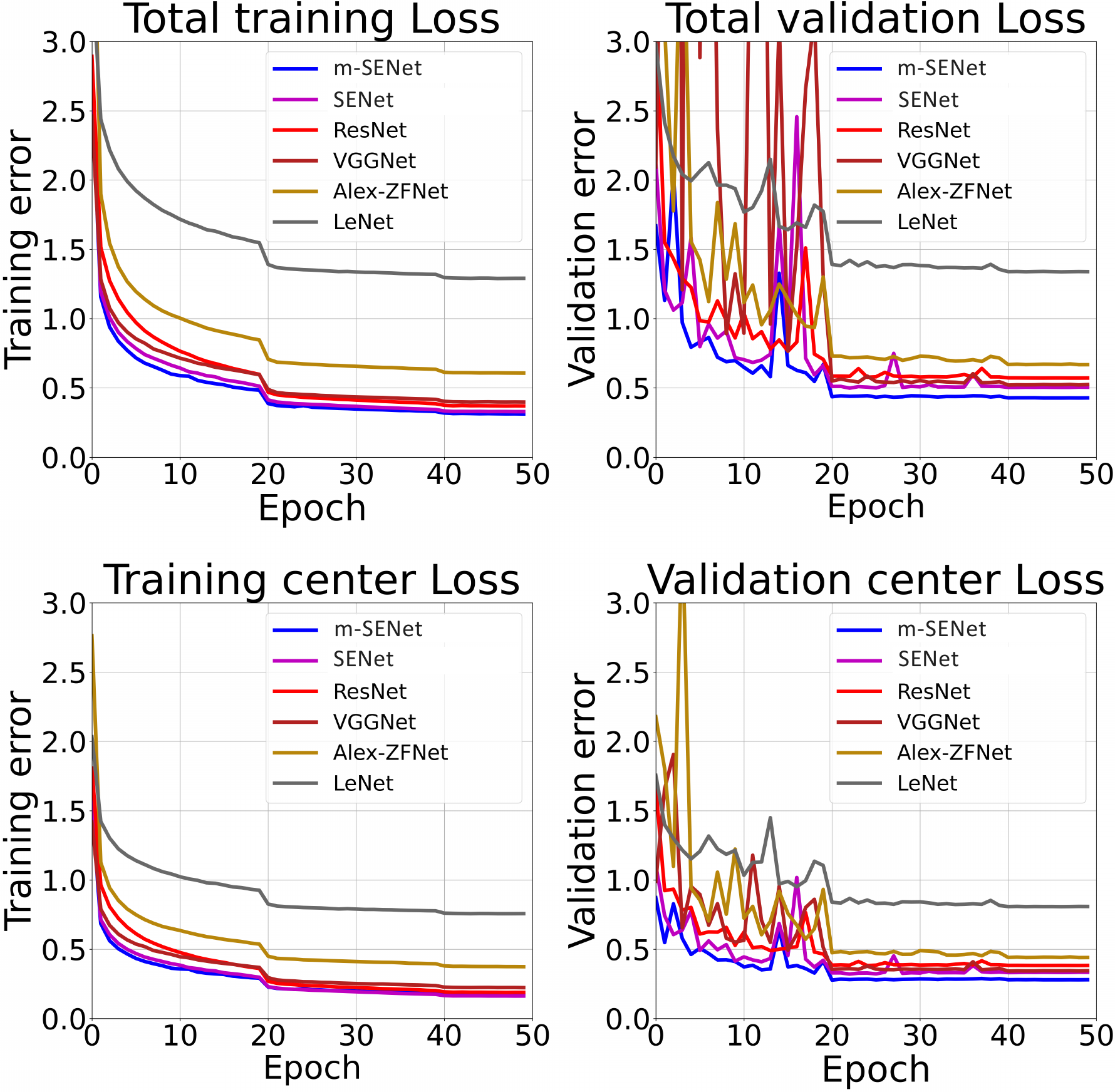}
  \caption{
  Total (upper row) and peak center loss functions (lower row) of six different CNN architectures, with respect to training epochs. Note that learning rate was reduced to 10\% after 20th and 40th epochs. Mean-square error loss function was employed. Left and right columns show training and validation losses, respectively. 
  }
  \label{fig:training}
\end{figure}

Fig.~\ref{fig:training} shows a summary of training and validation mean-square losses from different CNN architectures (SGD as optimizer). Among the four target properties (number of peaks in each spectrum and center, width, and amplitude of maximum area peak) peak center loss, which tends to yield the largest error, were plotted as well. A gradual reduction of both the training and validation loss as we go from LeNet to m-SENet is evident, showing better representational power of our m-SENet model in peak detection and fitting problem compared to other CNN architectures. 

\begin{figure}
  \centering
  \includegraphics[width=0.80\textwidth]{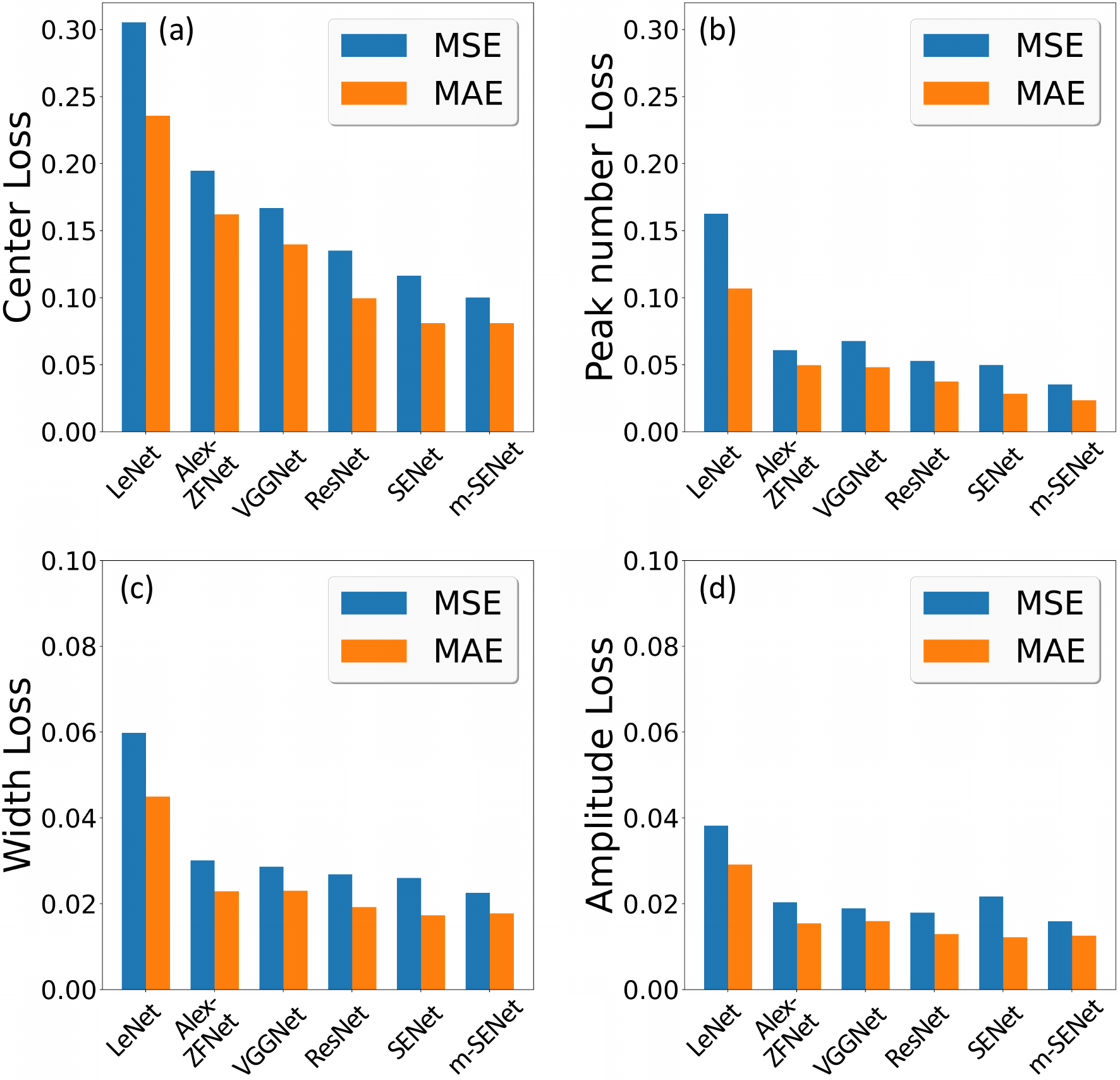}
  \caption{
  MAE loss functions ---  (a) center, (b) number of peaks, (c) width, and (d) amplitude of the largest-area peak --- of six different CNN architectures on the same test dataset. Blue and orange columns show results employing MSE and MAE losses during the network training and validation, respectively.
  }
  \label{fig:mse_vs_mae}
\end{figure}

Note that training and validation results shown in Fig.~\ref{fig:training} were obtained using mean-square-error (MSE) as the loss function, which is a widely adopted choice. In several cases, however, MSE tends to overestimate the effects of very few outliers in our randomly generated dataset and may lead to poor predicting performance of the model. In such cases, mean-absolute-error (MAE) can be a better option. Interestingly, our multiple peak fitting problem was found to be such one. 
To compare effects of MSE and MAE loss functions in model training and validation, we separately trained each CNN model using MSE and MAE, then applied each trained model in fitting the test dataset and computed MAE losses for comparison in equal footing.
Fig.~\ref{fig:mse_vs_mae} summarizes the results, where test losses of all four target variables (number of peaks, and center, width, and amplitude of the largest-area peak) in all CNN architectures show lower value when MAE was used in model training. From the result, it can be seen that SENet and m-SENet with MAE training loss function show almost the same fitting performance in this study. 

Once we confirmed that m-SENet shows the best fitting performance, we increased the size of the total dataset from 1.5 million to 10 million to enhance the performance of the model, because accurate peak fitting is crucial in the sequential subtraction of peaks from the spectrum. Table~\ref{tab:10m} compares loss function results from m-SENet trained with 1.5 and 10 million datasets, which shows clear improvements employing a large training dataset. Note that we haven't tested saturation of model performance as a function of the size of the training dataset due to the limitation of our computational resources. 

\begin{table}
\centering
\caption{Comparison between MAE loss functions from m-SENet trained with 1.5 and 10 millions datasets, where the values were obtained by applying same test dataset to both trained models.}
\vspace{.1in}
\begin{tabular}{lcccc}
\hline \hline
& center & width & amplitude & number of peaks \\
\hline 
1.5M set & 0.083 & 0.018 & 0.013 & 0.023 \\
10M set & 0.056 & 0.012 & 0.009 & 0.013 \\ \hline\hline
\end{tabular}
\label{tab:10m}
\end{table}

\subsection{Iterative peak subtraction via trained CNN and basin-hopping optimization}

\begin{figure}
  \centering
  \includegraphics[width=0.70\textwidth]{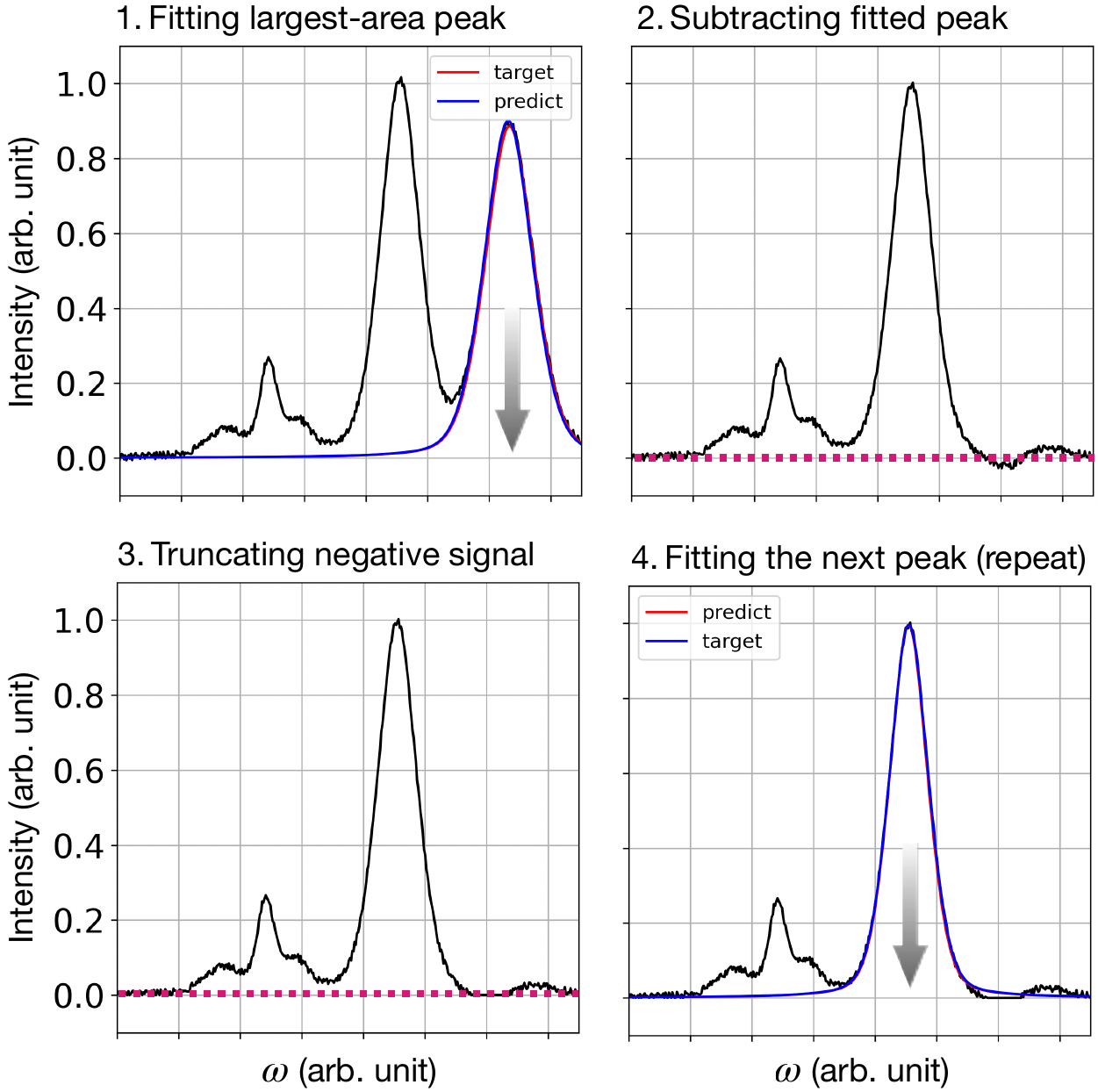}
  \caption{
  An illustration of iterative peak fitting and subtraction scheme we employed in this study. In the upper left and lower right panels, red and blue curves depict CNN-fitted and exact peak features respectively. Negative intensity caused by subtraction of inaccurate peaks (upper right panel, below the red dashed line) is truncated before applying subsequent peak fitting and subtraction.
  }
  \label{fig:iteration}
\end{figure}

After an accurate CNN model is prepared, it can be applied to sequentially subtract peak features from any given spectrum. Fig.~\ref{fig:iteration} illustrates our iterative peak fitting and subtraction scheme. A potential issue in this type of feature subtraction is, even very small errors in peak fitting may cause wiggly and negative features in the spectrum after subtraction which may be mistaken as some meaningful peaks by the CNN. This issue can be circumvented by i) accurately determining the number of total peaks at the initial fitting and performing subtractions exactly the number of times and ii) truncating negative intensity after subtraction (see lower left panel of Fig.~\ref{fig:iteration}). Because the accuracy of peak number prediction is quite high (check Fig.~\ref{fig:mse_vs_mae} and Table~\ref{tab:10m}), this should in most cases prevent the overall workflow from obtaining additional spurious peaks.

\begin{figure}
  \centering
  \includegraphics[width=0.70\textwidth]{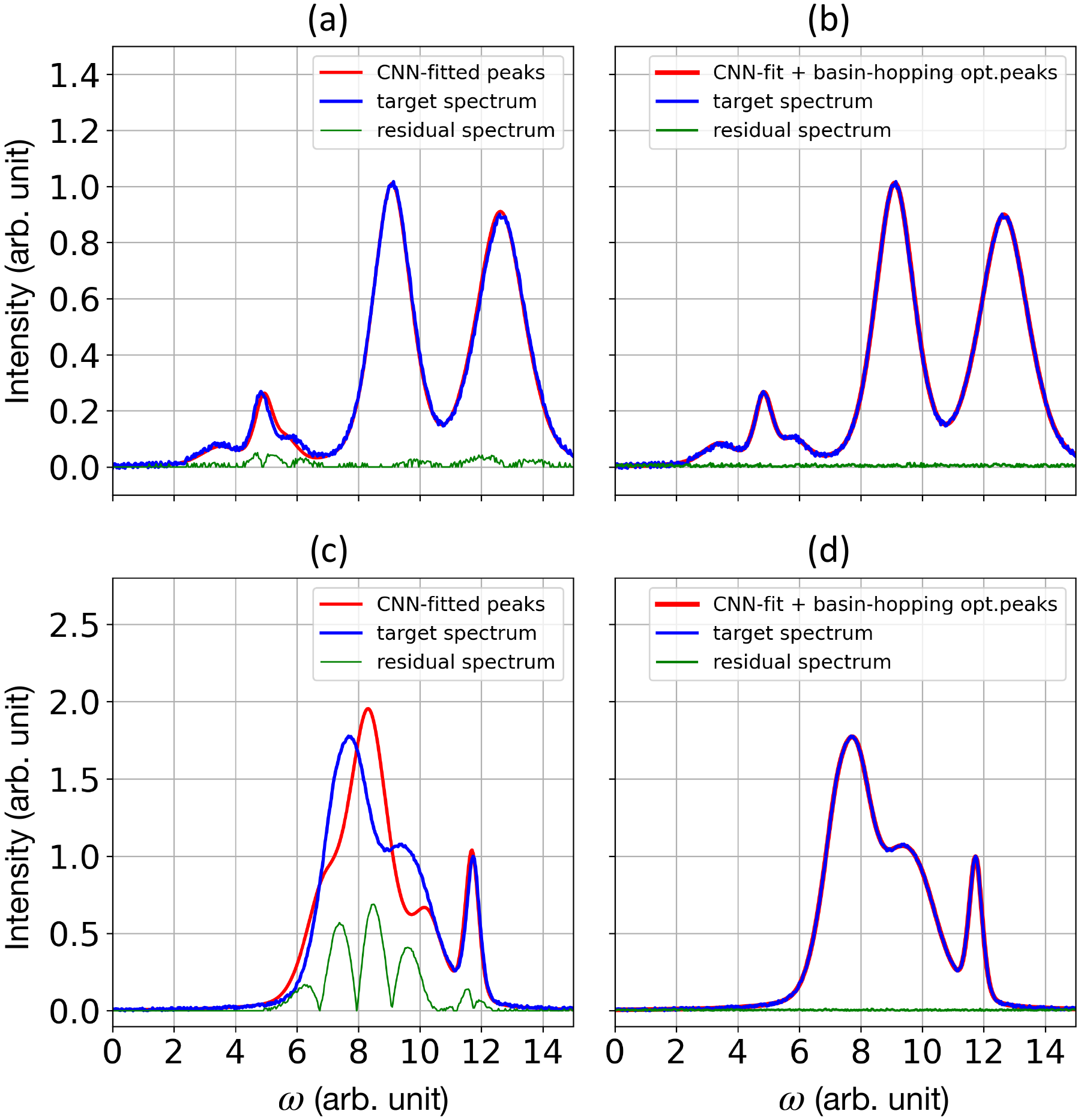}
  \caption{
  Comparison between peak fitting results before and after applying basin-hopping algorithm. Top (a,b) and bottom (c,d) rows show two different cases where the initial iterative CNN-fitting yield good and relative poor agreements with the target spectra, respectively, after which application of basin-hopping algorithm reduces any residual errors to almost zero. 
  }
  \label{fig:opt}
\end{figure}

Lastly, after extracting all peak features in a given spectrum, conventional global optimization methods such as basin-hopping algorithms\cite{basin-hopping} can be applied to further optimize fitting results and reduce any residual error to minimum. Basin-hopping can be considered an improved simulated annealing method, which involves in an additional local minimization stage before the accept/reject decision at each iteration\cite{basin-hopping}. We used \verb scipy.optimize.basinhopping  function implemented in the {\sc scipy} package\footnote{https://www.scipy.org}. Fig.~\ref{fig:opt} show an example optimization result on a couple of test data, where in Fig.~\ref{fig:opt}(a) the iterative CNN application already resulted in fairly good agreement with the target spectrum. In another case (Fig.~\ref{fig:opt}(b)), where CNN yields relatively poor agreement with the target spectrum, it can be checked that basin-hopping algorithm remedies the situation starting from the initial guess of peak parameters provided by the CNN model.

\section{Application to experimental photoemission spectra}

\begin{turnpage}
\begin{figure}
  \centering
  \includegraphics[width=1.40\textwidth]{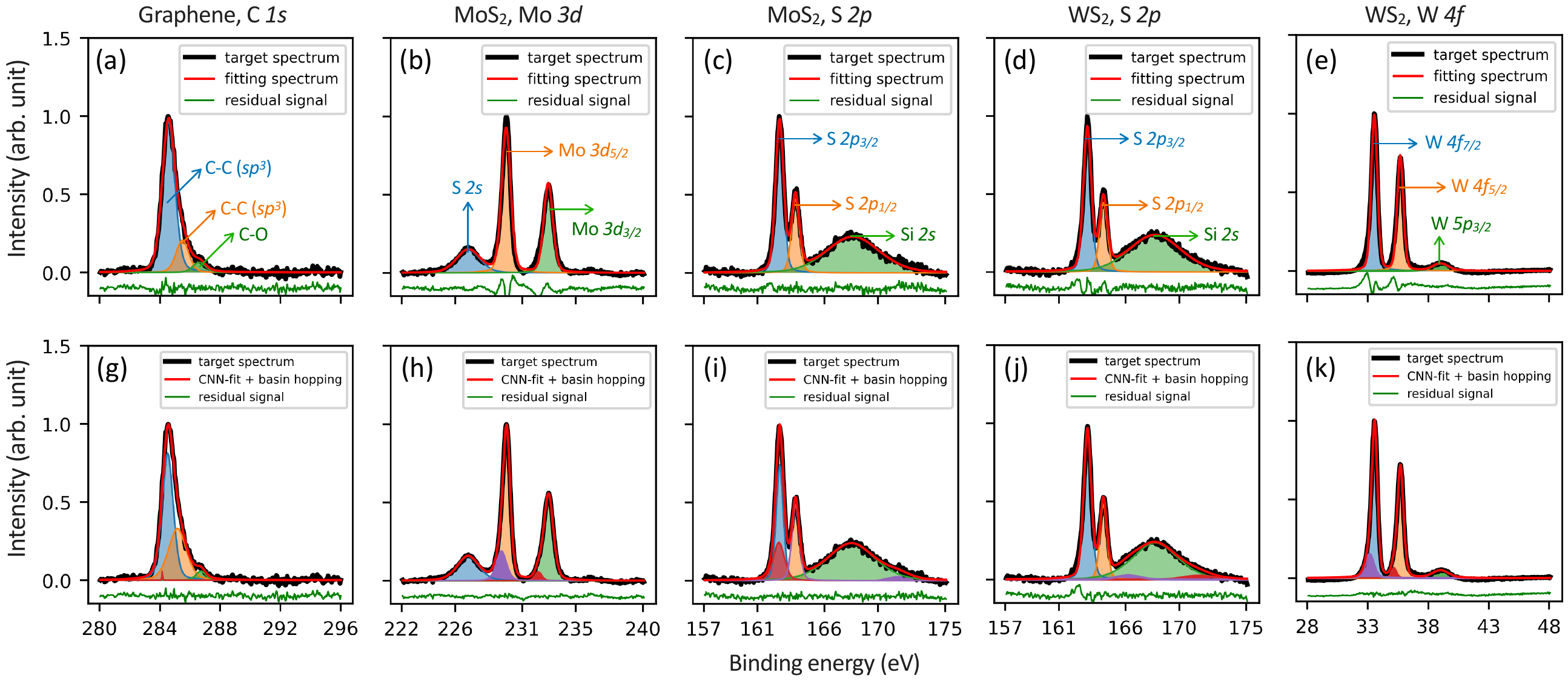}
  \caption{
  Human- (upper row) and machine-fitted (lower row) photoemission spectra of (a) graphene C $1s$, (b) MoS$_2$ Mo $3d$, (c) MoS$_2$ S $2p$, (d) WS$_2$ S $2p$, and (e) WS$_2$ W $4f$. In each panel black, red, and green curves represent raw data with background signal subtracted, fitted spectrum, and residual signal, respectively, where fitted peaks are depicted as colored filled curves. 
  }
  \label{fig:exp}
\end{figure}
\end{turnpage}

\begin{table}
\centering
\caption{Comparison of human-fitted and machine-fitted peak parameters from experimental photoemission spectra as shown in Fig.~{\ref{fig:exp}}.}
\vspace{.1in}
{\footnotesize
\begin{tabular}{l | ccc | ccc }
\hline \hline
&& Human-fitted &&& Machine-fitted & \\
& center (eV) & width (eV) & Norm. amp. & center (eV) & width (eV) & Norm. amp. \\
\hline 
Graphene, C $1s$ &&&&&& \\
Peak 1 & 284.60 & 1.00 & 0.96 & 284.50 & 1.08 & 0.81 \\
Peak 2 & 285.50 & 1.00 & 0.21 & 285.15 & 1.72 & 0.33 \\
Peak 3 & 286.60 & 1.00 & 0.06 & 286.71 & 0.97 & 0.06 \\
Peak 4 & -          &  -      & -       & 284.16 & 0.08 & 0.06 \\
\hline
MoS$_2$ on Si, Mo 3$d$ &&&&&& \\
Peak 1 & 226.98 & 2.06 & 0.15 & 226.94 & 2.08 & 0.16 \\
Peak 2 & 229.78 & 0.80 & 0.92 & 229.86 & 0.68 & 0.88 \\
Peak 3 & 232.93 & 0.80 & 0.57 & 232.97 & 0.92 & 0.55 \\
Peak 4 & -          &  -      & -       & 229.46 & 0.99 & 0.19 \\
Peak 5 & -          &  -      & -       & 232.20 & 0.68 & 0.06 \\
\hline
MoS$_2$ on Si, S 2$p$ &&&&&& \\
Peak 1 & 162.68 & 0.70 & 0.97 & 162.66 & 0.69 & 0.74 \\
Peak 2 & 163.83 & 0.70 & 0.47 & 163.83 & 0.72 & 0.49 \\
Peak 3 & 168.13 & 4.56 & 0.23 & 168.01 & 4.56 & 0.24 \\
Peak 4 & -          &  -      & -       & 162.61 & 1.08 & 0.24 \\
Peak 5 & -          & -       & -       & 171.52 & 2.34 & 0.03 \\
\hline
WS$_2$ on Si, S 2$p$ &&&&&& \\
Peak 1 & 163.13 & 0.70 & 0.91 & 163.15 & 0.74 & 0.95 \\
Peak 2 & 164.33 & 0.70 & 0.44 & 164.32 & 0.68 & 0.48 \\
Peak 3 & 168.18 & 4.46 & 0.24 & 168.19 & 4.35 & 0.24 \\
Peak 4 & -          &  -      & -       & 166.21 & 2.21 & 0.03 \\
Peak 5 & -          &  -      & -       & 171.48 & 2.78 & 0.03 \\
\hline 
WS$_2$ on Si, W 4$f$ &&&&&& \\
Peak 1 & 33.53 & 0.60 & 1.00 & 33.58 & 0.53 & 0.93 \\
Peak 2 & 35.68 & 0.60 & 0.73 & 35.68 & 0.60 & 0.71 \\
Peak 3 & 39.03 & 1.52 & 0.05 & 39.03 & 1.57 & 0.05 \\
Peak 4 & -         &  -      & -       & 33.13 & 0.82 & 0.15 \\
Peak 5 & -         &  -      & -       & 35.18 & 0.52 & 0.70 \\
\hline
\end{tabular}
}
\label{tab:exp}
\end{table}

Finally, we apply our fitting method to experimental x-ray photoemission spectra of graphene, MoS$_2$, and WS$_2$, and the results are summarized in Fig.~\ref{fig:exp}, where the top (Fig.~\ref{fig:exp}(a-e)) and bottom (Fig.~\ref{fig:exp}(g-k)) rows show human- and machine-fitted results, respectively. Note that background signals were subtracted from the raw data via \textsc{igor pro} package, and that defect-induced shoulder-like features in Mo $3d$ (orange- and green-shaded area in Fig.~\ref{fig:exp}(b)), Si $2s$ (green-shaded area in Fig.~\ref{fig:exp}(c,d)), and W $4f$ (blue- and orange-shades in Fig.~\ref{fig:exp}(e)) were intentionally ignored for the comparison with our machine-fitted results (see the residual signals therein, depicted as green curves). 

It can be seen that, in all cases, our machine-fitted results yield better agreement with the raw data compared to human-fitted results ({\it i.e.} smaller residual signal) and our previous result\cite{HPark2020}. Note that in the human-fitted results, fitting for Mo $3d$, Si $2s$, and W $4f$ peaks result in noticeable residual signals, signifying the defected-induced shoulder-like features. Indeed our iterative fitting model predicts the presence of additional peaks in those cases (darker red and violet shades in Fig.~\ref{fig:exp}(h,j,k), probably capturing overlooked signatures in human-fitting. 

On the other hand, note that our fitting model generates spurious additional peaks in C-C $sp^3$ in graphene and S $2p$ in MoS$_2$ (red shades in Fig.~\ref{fig:exp}(g) and (i), respectively), which may be attributed to the choice of pseudo-Voigt function with fixed internal parameters in Eq.~\ref{eq:psVoigt}. The use of true Voigt function with variable Lorentzian-Gaussian ratio to generate training dataset may resolve this issue.


\section{Discussion and conclusion}

We comment that, in the generation of the training dataset, we did not incorporate knowledge on physical processes that generate broadening of peaks, and considered peak width as a uniform random parameter in a range of our choice. Since excitations located close in energy should undergo similar scattering processes, most of the experimental photoemission peaks in a given energy range often show similar peak widths. Consideration of such observation in data generation may enhance the fitting performance of our model. Also, the employment of the true Voigt function instead of the pseudo-Voigt function that we used may improve the performance even further, which may reduce the occurrence of spurious peaks in the fitting. 

Overall, in this study, we've explored the potential power of CNN in analyzing spectroscopy data in the frequency domain and achieved significant improvement in fitting performance compared to our previous work\cite{HPark2020} via the employment of advanced deep CNN architectures and the application of iterative peak subtraction. We also benchmarked the fitting performance of various deep CNN networks and confirmed that the m-SENet that we propose in this study yields the best performance. With the help of conventional global optimization techniques, our method produces comparable or sometimes even better fitting performance compared to human fitting, and may potentially be applicable to high-throughput analyses of spectroscopic data with minimal human intervention. 

\section{Acknowledgements}
HSK acknowledges the support of the National Research Foundation of Korea (Basic Science Research Program, Grant No. 2020R1C1C1005900), and also thanks Hunpyo Lee. Juyong Lee, Hongkee Yoon, and Sooran Kim for fruitful discussions and insightful comments on the manuscript.

\bibliography{Peak_Fitting_JKPS.bib}

\begin{thebibliography}{34}%
\makeatletter
\providecommand \@ifxundefined [1]{%
 \@ifx{#1\undefined}
}%
\providecommand \@ifnum [1]{%
 \ifnum #1\expandafter \@firstoftwo
 \else \expandafter \@secondoftwo
 \fi
}%
\providecommand \@ifx [1]{%
 \ifx #1\expandafter \@firstoftwo
 \else \expandafter \@secondoftwo
 \fi
}%
\providecommand \natexlab [1]{#1}%
\providecommand \enquote  [1]{``#1''}%
\providecommand \bibnamefont  [1]{#1}%
\providecommand \bibfnamefont [1]{#1}%
\providecommand \citenamefont [1]{#1}%
\providecommand \href@noop [0]{\@secondoftwo}%
\providecommand \href [0]{\begingroup \@sanitize@url \@href}%
\providecommand \@href[1]{\@@startlink{#1}\@@href}%
\providecommand \@@href[1]{\endgroup#1\@@endlink}%
\providecommand \@sanitize@url [0]{\catcode `\\12\catcode `\$12\catcode
  `\&12\catcode `\#12\catcode `\^12\catcode `\_12\catcode `\%12\relax}%
\providecommand \@@startlink[1]{}%
\providecommand \@@endlink[0]{}%
\providecommand \url  [0]{\begingroup\@sanitize@url \@url }%
\providecommand \@url [1]{\endgroup\@href {#1}{\urlprefix }}%
\providecommand \urlprefix  [0]{URL }%
\providecommand \Eprint [0]{\href }%
\providecommand \doibase [0]{http://dx.doi.org/}%
\providecommand \selectlanguage [0]{\@gobble}%
\providecommand \bibinfo  [0]{\@secondoftwo}%
\providecommand \bibfield  [0]{\@secondoftwo}%
\providecommand \translation [1]{[#1]}%
\providecommand \BibitemOpen [0]{}%
\providecommand \bibitemStop [0]{}%
\providecommand \bibitemNoStop [0]{.\EOS\space}%
\providecommand \EOS [0]{\spacefactor3000\relax}%
\providecommand \BibitemShut  [1]{\csname bibitem#1\endcsname}%
\let\auto@bib@innerbib\@empty
\bibitem [{\citenamefont {Parr}(1983)}]{DFT_Review1}%
  \BibitemOpen
  \bibfield  {author} {\bibinfo {author} {\bibfnamefont {R~G}\ \bibnamefont
  {Parr}},\ }\bibfield  {title} {\enquote {\bibinfo {title} {Density functional
  theory},}\ }\href {\doibase 10.1146/annurev.pc.34.100183.003215} {\bibfield
  {journal} {\bibinfo  {journal} {Ann. Rev. Phys. Chem.}\ }\textbf {\bibinfo
  {volume} {34}},\ \bibinfo {pages} {631--656} (\bibinfo {year} {1983})},\
  \Eprint
  {http://arxiv.org/abs/https://doi.org/10.1146/annurev.pc.34.100183.003215}
  {https://doi.org/10.1146/annurev.pc.34.100183.003215} \BibitemShut {NoStop}%
\bibitem [{\citenamefont {Burke}(2012)}]{DFT_Review2}%
  \BibitemOpen
  \bibfield  {author} {\bibinfo {author} {\bibfnamefont {Kieron}\ \bibnamefont
  {Burke}},\ }\bibfield  {title} {\enquote {\bibinfo {title} {Perspective on
  density functional theory},}\ }\href {\doibase 10.1063/1.4704546} {\bibfield
  {journal} {\bibinfo  {journal} {J. Chem. Phys.}\ }\textbf {\bibinfo {volume}
  {136}},\ \bibinfo {pages} {150901} (\bibinfo {year} {2012})},\ \Eprint
  {http://arxiv.org/abs/https://doi.org/10.1063/1.4704546}
  {https://doi.org/10.1063/1.4704546} \BibitemShut {NoStop}%
\bibitem [{Note1()}]{Note1}%
  \BibitemOpen
  \bibinfo {note} {Https://materialsproject.org}\BibitemShut {NoStop}%
\bibitem [{Note2()}]{Note2}%
  \BibitemOpen
  \bibinfo {note} {Http://oqmd.org}\BibitemShut {NoStop}%
\bibitem [{Note3()}]{Note3}%
  \BibitemOpen
  \bibinfo {note} {Https://nomad-coe.eu}\BibitemShut {NoStop}%
\bibitem [{\citenamefont {{Karpatne}}\ \emph {et~al.}(2017)\citenamefont
  {{Karpatne}}, \citenamefont {{Atluri}}, \citenamefont {{Faghmous}},
  \citenamefont {{Steinbach}}, \citenamefont {{Banerjee}}, \citenamefont
  {{Ganguly}}, \citenamefont {{Shekhar}}, \citenamefont {{Samatova}},\ and\
  \citenamefont {{Kumar}}}]{datascience2017}%
  \BibitemOpen
  \bibfield  {author} {\bibinfo {author} {\bibfnamefont {A.}~\bibnamefont
  {{Karpatne}}}, \bibinfo {author} {\bibfnamefont {G.}~\bibnamefont
  {{Atluri}}}, \bibinfo {author} {\bibfnamefont {J.~H.}\ \bibnamefont
  {{Faghmous}}}, \bibinfo {author} {\bibfnamefont {M.}~\bibnamefont
  {{Steinbach}}}, \bibinfo {author} {\bibfnamefont {A.}~\bibnamefont
  {{Banerjee}}}, \bibinfo {author} {\bibfnamefont {A.}~\bibnamefont
  {{Ganguly}}}, \bibinfo {author} {\bibfnamefont {S.}~\bibnamefont
  {{Shekhar}}}, \bibinfo {author} {\bibfnamefont {N.}~\bibnamefont
  {{Samatova}}}, \ and\ \bibinfo {author} {\bibfnamefont {V.}~\bibnamefont
  {{Kumar}}},\ }\bibfield  {title} {\enquote {\bibinfo {title} {Theory-guided
  data science: A new paradigm for scientific discovery from data},}\
  }\href@noop {} {\bibfield  {journal} {\bibinfo  {journal} {IEEE Trans. Knowl.
  Data. Eng.}\ }\textbf {\bibinfo {volume} {29}},\ \bibinfo {pages}
  {2318--2331} (\bibinfo {year} {2017})}\BibitemShut {NoStop}%
\bibitem [{\citenamefont {Xie}\ and\ \citenamefont
  {Grossman}(2018)}]{CGCNN2018}%
  \BibitemOpen
  \bibfield  {author} {\bibinfo {author} {\bibfnamefont {Tian}\ \bibnamefont
  {Xie}}\ and\ \bibinfo {author} {\bibfnamefont {Jeffrey~C.}\ \bibnamefont
  {Grossman}},\ }\bibfield  {title} {\enquote {\bibinfo {title} {Crystal graph
  convolutional neural networks for an accurate and interpretable prediction of
  material properties},}\ }\href@noop {} {\bibfield  {journal} {\bibinfo
  {journal} {Phys. Rev. Lett.}\ }\textbf {\bibinfo {volume} {120}},\ \bibinfo
  {pages} {145301} (\bibinfo {year} {2018})}\BibitemShut {NoStop}%
\bibitem [{\citenamefont {Butler}\ \emph {et~al.}(2018)\citenamefont {Butler},
  \citenamefont {Davies}, \citenamefont {Cartwright}, \citenamefont {Isayev},\
  and\ \citenamefont {Walsh}}]{Butler2018}%
  \BibitemOpen
  \bibfield  {author} {\bibinfo {author} {\bibfnamefont {Keith~T.}\
  \bibnamefont {Butler}}, \bibinfo {author} {\bibfnamefont {Daniel~W.}\
  \bibnamefont {Davies}}, \bibinfo {author} {\bibfnamefont {Hugh}\ \bibnamefont
  {Cartwright}}, \bibinfo {author} {\bibfnamefont {Olexandr}\ \bibnamefont
  {Isayev}}, \ and\ \bibinfo {author} {\bibfnamefont {Aron}\ \bibnamefont
  {Walsh}},\ }\bibfield  {title} {\enquote {\bibinfo {title} {Machine learning
  for molecular and materials science},}\ }\href {\doibase
  10.1038/s41586-018-0337-2} {\bibfield  {journal} {\bibinfo  {journal}
  {Nature}\ }\textbf {\bibinfo {volume} {559}},\ \bibinfo {pages} {547--555}
  (\bibinfo {year} {2018})}\BibitemShut {NoStop}%
\bibitem [{\citenamefont {Zunger}(2018)}]{Zunger2018}%
  \BibitemOpen
  \bibfield  {author} {\bibinfo {author} {\bibfnamefont {Alex}\ \bibnamefont
  {Zunger}},\ }\bibfield  {title} {\enquote {\bibinfo {title} {Inverse design
  in search of materials with target functionalities},}\ }\href {\doibase
  10.1038/s41570-018-0121} {\bibfield  {journal} {\bibinfo  {journal} {Nat.
  Rev. Chem.}\ }\textbf {\bibinfo {volume} {2}},\ \bibinfo {pages} {0121}
  (\bibinfo {year} {2018})}\BibitemShut {NoStop}%
\bibitem [{\citenamefont {Sanchez-Lengeling}\ and\ \citenamefont
  {Aspuru-Guzik}(2018)}]{Sanchez-Lengeling360}%
  \BibitemOpen
  \bibfield  {author} {\bibinfo {author} {\bibfnamefont {Benjamin}\
  \bibnamefont {Sanchez-Lengeling}}\ and\ \bibinfo {author} {\bibfnamefont
  {Al{\'a}n}\ \bibnamefont {Aspuru-Guzik}},\ }\bibfield  {title} {\enquote
  {\bibinfo {title} {Inverse molecular design using machine learning:
  Generative models for matter engineering},}\ }\href {\doibase
  10.1126/science.aat2663} {\bibfield  {journal} {\bibinfo  {journal}
  {Science}\ }\textbf {\bibinfo {volume} {361}},\ \bibinfo {pages} {360--365}
  (\bibinfo {year} {2018})}\BibitemShut {NoStop}%
\bibitem [{\citenamefont {Schleder}\ \emph {et~al.}(2019)\citenamefont
  {Schleder}, \citenamefont {Padilha}, \citenamefont {Acosta}, \citenamefont
  {Costa},\ and\ \citenamefont {Fazzio}}]{Schleder_2019}%
  \BibitemOpen
  \bibfield  {author} {\bibinfo {author} {\bibfnamefont {Gabriel~R}\
  \bibnamefont {Schleder}}, \bibinfo {author} {\bibfnamefont {Antonio C~M}\
  \bibnamefont {Padilha}}, \bibinfo {author} {\bibfnamefont {Carlos~Mera}\
  \bibnamefont {Acosta}}, \bibinfo {author} {\bibfnamefont {Marcio}\
  \bibnamefont {Costa}}, \ and\ \bibinfo {author} {\bibfnamefont {Adalberto}\
  \bibnamefont {Fazzio}},\ }\bibfield  {title} {\enquote {\bibinfo {title}
  {From {DFT} to machine learning: recent approaches to materials
  science{\textendash}a review},}\ }\href {\doibase 10.1088/2515-7639/ab084b}
  {\bibfield  {journal} {\bibinfo  {journal} {J. Phys.: Mater.}\ }\textbf
  {\bibinfo {volume} {2}},\ \bibinfo {pages} {032001} (\bibinfo {year}
  {2019})}\BibitemShut {NoStop}%
\bibitem [{\citenamefont {Himanen}\ \emph {et~al.}(2019)\citenamefont
  {Himanen}, \citenamefont {Geurts}, \citenamefont {Foster},\ and\
  \citenamefont {Rinke}}]{Himanen_2019}%
  \BibitemOpen
  \bibfield  {author} {\bibinfo {author} {\bibfnamefont {Lauri}\ \bibnamefont
  {Himanen}}, \bibinfo {author} {\bibfnamefont {Amber}\ \bibnamefont {Geurts}},
  \bibinfo {author} {\bibfnamefont {Adam~Stuart}\ \bibnamefont {Foster}}, \
  and\ \bibinfo {author} {\bibfnamefont {Patrick}\ \bibnamefont {Rinke}},\
  }\bibfield  {title} {\enquote {\bibinfo {title} {Data-driven materials
  science: Status, challenges, and perspectives},}\ }\href {\doibase
  https://doi.org/10.1002/advs.201900808} {\bibfield  {journal} {\bibinfo
  {journal} {Adv. Sci.}\ }\textbf {\bibinfo {volume} {6}},\ \bibinfo {pages}
  {1900808} (\bibinfo {year} {2019})},\ \Eprint
  {http://arxiv.org/abs/https://onlinelibrary.wiley.com/doi/pdf/10.1002/advs.201900808}
  {https://onlinelibrary.wiley.com/doi/pdf/10.1002/advs.201900808} \BibitemShut
  {NoStop}%
\bibitem [{\citenamefont {Zhang}\ \emph {et~al.}(2019)\citenamefont {Zhang},
  \citenamefont {Mesaros}, \citenamefont {Fujita}, \citenamefont {Edkins},
  \citenamefont {Hamidian}, \citenamefont {Ch'ng}, \citenamefont {Eisaki},
  \citenamefont {Uchida}, \citenamefont {Davis}, \citenamefont {Khatami},\ and\
  \citenamefont {Kim}}]{Zhang2019}%
  \BibitemOpen
  \bibfield  {author} {\bibinfo {author} {\bibfnamefont {Yi}~\bibnamefont
  {Zhang}}, \bibinfo {author} {\bibfnamefont {A.}~\bibnamefont {Mesaros}},
  \bibinfo {author} {\bibfnamefont {K.}~\bibnamefont {Fujita}}, \bibinfo
  {author} {\bibfnamefont {S.~D.}\ \bibnamefont {Edkins}}, \bibinfo {author}
  {\bibfnamefont {M.~H.}\ \bibnamefont {Hamidian}}, \bibinfo {author}
  {\bibfnamefont {K.}~\bibnamefont {Ch'ng}}, \bibinfo {author} {\bibfnamefont
  {H.}~\bibnamefont {Eisaki}}, \bibinfo {author} {\bibfnamefont
  {S.}~\bibnamefont {Uchida}}, \bibinfo {author} {\bibfnamefont
  {J.~C.~S{\'e}amus}\ \bibnamefont {Davis}}, \bibinfo {author} {\bibfnamefont
  {Ehsan}\ \bibnamefont {Khatami}}, \ and\ \bibinfo {author} {\bibfnamefont
  {Eun-Ah}\ \bibnamefont {Kim}},\ }\bibfield  {title} {\enquote {\bibinfo
  {title} {Machine learning in electronic-quantum-matter imaging
  experiments},}\ }\href {\doibase 10.1038/s41586-019-1319-8} {\bibfield
  {journal} {\bibinfo  {journal} {Nature}\ }\textbf {\bibinfo {volume} {570}},\
  \bibinfo {pages} {484--490} (\bibinfo {year} {2019})}\BibitemShut {NoStop}%
\bibitem [{\citenamefont {Drera}\ \emph {et~al.}(2020)\citenamefont {Drera},
  \citenamefont {Kropf},\ and\ \citenamefont {Sangaletti}}]{Drera_2020}%
  \BibitemOpen
  \bibfield  {author} {\bibinfo {author} {\bibfnamefont {G}~\bibnamefont
  {Drera}}, \bibinfo {author} {\bibfnamefont {C~M}\ \bibnamefont {Kropf}}, \
  and\ \bibinfo {author} {\bibfnamefont {L}~\bibnamefont {Sangaletti}},\
  }\bibfield  {title} {\enquote {\bibinfo {title} {Deep neural network for
  x-ray photoelectron spectroscopy data analysis},}\ }\href {\doibase
  10.1088/2632-2153/ab5da6} {\bibfield  {journal} {\bibinfo  {journal} {Mach.
  Learn.: Sci. Technol.}\ }\textbf {\bibinfo {volume} {1}},\ \bibinfo {pages}
  {015008} (\bibinfo {year} {2020})}\BibitemShut {NoStop}%
\bibitem [{\citenamefont {Yamaji}\ \emph {et~al.}(2019)\citenamefont {Yamaji},
  \citenamefont {Yoshida}, \citenamefont {Fujimori},\ and\ \citenamefont
  {Imada}}]{Youhei_2019}%
  \BibitemOpen
  \bibfield  {author} {\bibinfo {author} {\bibfnamefont {Youhei}\ \bibnamefont
  {Yamaji}}, \bibinfo {author} {\bibfnamefont {Teppei}\ \bibnamefont
  {Yoshida}}, \bibinfo {author} {\bibfnamefont {Atsushi}\ \bibnamefont
  {Fujimori}}, \ and\ \bibinfo {author} {\bibfnamefont {Masatoshi}\
  \bibnamefont {Imada}},\ }\href {http://arxiv.org/abs/1903.08060} {\enquote
  {\bibinfo {title} {Hidden self-energies as origin of cuprate
  superconductivity revealed by machine learning},}\ } (\bibinfo {year}
  {2019})\BibitemShut {NoStop}%
\bibitem [{\citenamefont {Ida}\ \emph {et~al.}(2000)\citenamefont {Ida},
  \citenamefont {Ando},\ and\ \citenamefont {Toraya}}]{Ida:nt0146}%
  \BibitemOpen
  \bibfield  {author} {\bibinfo {author} {\bibfnamefont {T.}~\bibnamefont
  {Ida}}, \bibinfo {author} {\bibfnamefont {M.}~\bibnamefont {Ando}}, \ and\
  \bibinfo {author} {\bibfnamefont {H.}~\bibnamefont {Toraya}},\ }\bibfield
  {title} {\enquote {\bibinfo {title} {{Extended pseudo-Voigt function for
  approximating the Voigt profile}},}\ }\href {\doibase
  10.1107/S0021889800010219} {\bibfield  {journal} {\bibinfo  {journal} {J.
  Appl. Crystallogr.}\ }\textbf {\bibinfo {volume} {33}},\ \bibinfo {pages}
  {1311--1316} (\bibinfo {year} {2000})}\BibitemShut {NoStop}%
\bibitem [{\citenamefont {LeCun}\ \emph {et~al.}(1998)\citenamefont {LeCun},
  \citenamefont {Bottou}, \citenamefont {Bengio},\ and\ \citenamefont
  {Haffner}}]{lecun1998gradient}%
  \BibitemOpen
  \bibfield  {author} {\bibinfo {author} {\bibfnamefont {Yann}\ \bibnamefont
  {LeCun}}, \bibinfo {author} {\bibfnamefont {L{\'e}on}\ \bibnamefont
  {Bottou}}, \bibinfo {author} {\bibfnamefont {Yoshua}\ \bibnamefont {Bengio}},
  \ and\ \bibinfo {author} {\bibfnamefont {Patrick}\ \bibnamefont {Haffner}},\
  }\bibfield  {title} {\enquote {\bibinfo {title} {Gradient-based learning
  applied to document recognition},}\ }\href@noop {} {\bibfield  {journal}
  {\bibinfo  {journal} {Proc. IEEE}\ }\textbf {\bibinfo {volume} {86}},\
  \bibinfo {pages} {2278--2324} (\bibinfo {year} {1998})}\BibitemShut {NoStop}%
\bibitem [{\citenamefont {LeCun}\ \emph {et~al.}(1989)\citenamefont {LeCun},
  \citenamefont {Boser}, \citenamefont {Denker}, \citenamefont {Henderson},
  \citenamefont {Howard}, \citenamefont {Hubbard},\ and\ \citenamefont
  {Jackel}}]{lecun1989backpropagation}%
  \BibitemOpen
  \bibfield  {author} {\bibinfo {author} {\bibfnamefont {Yann}\ \bibnamefont
  {LeCun}}, \bibinfo {author} {\bibfnamefont {Bernhard}\ \bibnamefont {Boser}},
  \bibinfo {author} {\bibfnamefont {John~S}\ \bibnamefont {Denker}}, \bibinfo
  {author} {\bibfnamefont {Donnie}\ \bibnamefont {Henderson}}, \bibinfo
  {author} {\bibfnamefont {Richard~E}\ \bibnamefont {Howard}}, \bibinfo
  {author} {\bibfnamefont {Wayne}\ \bibnamefont {Hubbard}}, \ and\ \bibinfo
  {author} {\bibfnamefont {Lawrence~D}\ \bibnamefont {Jackel}},\ }\bibfield
  {title} {\enquote {\bibinfo {title} {Backpropagation applied to handwritten
  zip code recognition},}\ }\href@noop {} {\bibfield  {journal} {\bibinfo
  {journal} {Neural Comput.}\ }\textbf {\bibinfo {volume} {1}},\ \bibinfo
  {pages} {541--551} (\bibinfo {year} {1989})}\BibitemShut {NoStop}%
\bibitem [{\citenamefont {Krizhevsky}\ \emph {et~al.}(2012)\citenamefont
  {Krizhevsky}, \citenamefont {Sutskever},\ and\ \citenamefont
  {Hinton}}]{krizhevsky2012imagenet}%
  \BibitemOpen
  \bibfield  {author} {\bibinfo {author} {\bibfnamefont {Alex}\ \bibnamefont
  {Krizhevsky}}, \bibinfo {author} {\bibfnamefont {Ilya}\ \bibnamefont
  {Sutskever}}, \ and\ \bibinfo {author} {\bibfnamefont {Geoffrey~E}\
  \bibnamefont {Hinton}},\ }\bibfield  {title} {\enquote {\bibinfo {title}
  {Imagenet classification with deep convolutional neural networks},}\
  }\href@noop {} {\bibfield  {journal} {\bibinfo  {journal} {Adv. Neural Inf.
  Process. Syst.}\ }\textbf {\bibinfo {volume} {25}},\ \bibinfo {pages}
  {1097--1105} (\bibinfo {year} {2012})}\BibitemShut {NoStop}%
\bibitem [{\citenamefont {Zeiler}\ and\ \citenamefont
  {Fergus}(2014)}]{zeiler2014visualizing}%
  \BibitemOpen
  \bibfield  {author} {\bibinfo {author} {\bibfnamefont {Matthew~D}\
  \bibnamefont {Zeiler}}\ and\ \bibinfo {author} {\bibfnamefont {Rob}\
  \bibnamefont {Fergus}},\ }\bibfield  {title} {\enquote {\bibinfo {title}
  {Visualizing and understanding convolutional networks},}\ }in\ \href@noop {}
  {\emph {\bibinfo {booktitle} {European conference on computer vision}}}\
  (\bibinfo {organization} {Springer},\ \bibinfo {year} {2014})\ pp.\ \bibinfo
  {pages} {818--833}\BibitemShut {NoStop}%
\bibitem [{\citenamefont {Simonyan}\ and\ \citenamefont
  {Zisserman}(2014)}]{simonyan2014very}%
  \BibitemOpen
  \bibfield  {author} {\bibinfo {author} {\bibfnamefont {Karen}\ \bibnamefont
  {Simonyan}}\ and\ \bibinfo {author} {\bibfnamefont {Andrew}\ \bibnamefont
  {Zisserman}},\ }\bibfield  {title} {\enquote {\bibinfo {title} {Very deep
  convolutional networks for large-scale image recognition},}\ }\href@noop {}
  {\bibfield  {journal} {\bibinfo  {journal} {arXiv preprint arXiv:1409.1556}\
  } (\bibinfo {year} {2014})}\BibitemShut {NoStop}%
\bibitem [{\citenamefont {He}\ \emph {et~al.}(2016)\citenamefont {He},
  \citenamefont {Zhang}, \citenamefont {Ren},\ and\ \citenamefont
  {Sun}}]{he2016deep}%
  \BibitemOpen
  \bibfield  {author} {\bibinfo {author} {\bibfnamefont {Kaiming}\ \bibnamefont
  {He}}, \bibinfo {author} {\bibfnamefont {Xiangyu}\ \bibnamefont {Zhang}},
  \bibinfo {author} {\bibfnamefont {Shaoqing}\ \bibnamefont {Ren}}, \ and\
  \bibinfo {author} {\bibfnamefont {Jian}\ \bibnamefont {Sun}},\ }\bibfield
  {title} {\enquote {\bibinfo {title} {Deep residual learning for image
  recognition},}\ }in\ \href@noop {} {\emph {\bibinfo {booktitle} {Proceedings
  of the IEEE conference on computer vision and pattern recognition}}}\
  (\bibinfo {year} {2016})\ pp.\ \bibinfo {pages} {770--778}\BibitemShut
  {NoStop}%
\bibitem [{\citenamefont {Hu}\ \emph {et~al.}(2018)\citenamefont {Hu},
  \citenamefont {Shen},\ and\ \citenamefont {Sun}}]{hu2018squeeze}%
  \BibitemOpen
  \bibfield  {author} {\bibinfo {author} {\bibfnamefont {Jie}\ \bibnamefont
  {Hu}}, \bibinfo {author} {\bibfnamefont {Li}~\bibnamefont {Shen}}, \ and\
  \bibinfo {author} {\bibfnamefont {Gang}\ \bibnamefont {Sun}},\ }\bibfield
  {title} {\enquote {\bibinfo {title} {Squeeze-and-excitation networks},}\ }in\
  \href@noop {} {\emph {\bibinfo {booktitle} {Proceedings of the IEEE
  conference on computer vision and pattern recognition}}}\ (\bibinfo {year}
  {2018})\ pp.\ \bibinfo {pages} {7132--7141}\BibitemShut {NoStop}%
\bibitem [{Note4()}]{Note4}%
  \BibitemOpen
  \bibinfo {note} {$\omega _0$, $I_0$, and $\delta $ are randomly chosen in the
  range of $0.13 \omega _{\protect \rm max} < \omega _0 < 0.87 \omega
  _{\protect \rm max}$, $0.02 \omega _{\protect \rm max} < \delta < 0.13 \omega
  _{\protect \rm max}$, and $0.05 < I_0 < 1.05$, respectively, where $0 \leq
  \omega \leq \omega _{\protect \rm max}$ is our frequency domain.}\BibitemShut
  {Stop}%
\bibitem [{\citenamefont {Kim}\ \emph {et~al.}(2018)\citenamefont {Kim},
  \citenamefont {Lee},\ and\ \citenamefont {Lee}}]{Kim_2018}%
  \BibitemOpen
  \bibfield  {author} {\bibinfo {author} {\bibfnamefont {Hyounggi}\
  \bibnamefont {Kim}}, \bibinfo {author} {\bibfnamefont {Hyunchan}\
  \bibnamefont {Lee}}, \ and\ \bibinfo {author} {\bibfnamefont {Hyunbok}\
  \bibnamefont {Lee}},\ }\bibfield  {title} {\enquote {\bibinfo {title}
  {Fabrication of poly(3-hexylthiophene-2,5-diyl) films with electrospray
  deposition method},}\ }\href {\doibase 10.7567/jjap.57.071601} {\bibfield
  {journal} {\bibinfo  {journal} {Jpn. J. Appl. Phys.}\ }\textbf {\bibinfo
  {volume} {57}},\ \bibinfo {pages} {071601} (\bibinfo {year}
  {2018})}\BibitemShut {NoStop}%
\bibitem [{\citenamefont {{Schmidt}}\ \emph {et~al.}(2019)\citenamefont
  {{Schmidt}}, \citenamefont {{Alstrøm}}, \citenamefont {{Svendstorp}},\ and\
  \citenamefont {{Larsen}}}]{PeakFitting}%
  \BibitemOpen
  \bibfield  {author} {\bibinfo {author} {\bibfnamefont {M.~N.}\ \bibnamefont
  {{Schmidt}}}, \bibinfo {author} {\bibfnamefont {T.~S.}\ \bibnamefont
  {{Alstrøm}}}, \bibinfo {author} {\bibfnamefont {M.}~\bibnamefont
  {{Svendstorp}}}, \ and\ \bibinfo {author} {\bibfnamefont {J.}~\bibnamefont
  {{Larsen}}},\ }\bibfield  {title} {\enquote {\bibinfo {title} {Peak detection
  and baseline correction using a convolutional neural network},}\ }in\
  \href@noop {} {\emph {\bibinfo {booktitle} {ICASSP 2019 - 2019 IEEE
  International Conference on Acoustics, Speech and Signal Processing
  (ICASSP)}}}\ (\bibinfo {year} {2019})\ pp.\ \bibinfo {pages}
  {2757--2761}\BibitemShut {NoStop}%
\bibitem [{\citenamefont {Chollet}\ \emph {et~al.}(2015)\citenamefont {Chollet}
  \emph {et~al.}}]{chollet2015keras}%
  \BibitemOpen
  \bibfield  {author} {\bibinfo {author} {\bibfnamefont {Fran\c{c}ois}\
  \bibnamefont {Chollet}} \emph {et~al.},\ }\href@noop {} {\enquote {\bibinfo
  {title} {Keras},}\ }\bibinfo {howpublished} {\url{https://keras.io}}
  (\bibinfo {year} {2015})\BibitemShut {NoStop}%
\bibitem [{\citenamefont {Abadi}\ \emph {et~al.}(2015)\citenamefont {Abadi},
  \citenamefont {Agarwal}, \citenamefont {Barham}, \citenamefont {Brevdo},
  \citenamefont {Chen}, \citenamefont {Citro}, \citenamefont {Corrado},
  \citenamefont {Davis}, \citenamefont {Dean}, \citenamefont {Devin},
  \citenamefont {Ghemawat}, \citenamefont {Goodfellow}, \citenamefont {Harp},
  \citenamefont {Irving}, \citenamefont {Isard}, \citenamefont {Jia},
  \citenamefont {Jozefowicz}, \citenamefont {Kaiser}, \citenamefont {Kudlur},
  \citenamefont {Levenberg}, \citenamefont {Man\'{e}}, \citenamefont {Monga},
  \citenamefont {Moore}, \citenamefont {Murray}, \citenamefont {Olah},
  \citenamefont {Schuster}, \citenamefont {Shlens}, \citenamefont {Steiner},
  \citenamefont {Sutskever}, \citenamefont {Talwar}, \citenamefont {Tucker},
  \citenamefont {Vanhoucke}, \citenamefont {Vasudevan}, \citenamefont
  {Vi\'{e}gas}, \citenamefont {Vinyals}, \citenamefont {Warden}, \citenamefont
  {Wattenberg}, \citenamefont {Wicke}, \citenamefont {Yu},\ and\ \citenamefont
  {Zheng}}]{tensorflow2015-whitepaper}%
  \BibitemOpen
  \bibfield  {author} {\bibinfo {author} {\bibfnamefont {Mart\'{\i}n}\
  \bibnamefont {Abadi}}, \bibinfo {author} {\bibfnamefont {Ashish}\
  \bibnamefont {Agarwal}}, \bibinfo {author} {\bibfnamefont {Paul}\
  \bibnamefont {Barham}}, \bibinfo {author} {\bibfnamefont {Eugene}\
  \bibnamefont {Brevdo}}, \bibinfo {author} {\bibfnamefont {Zhifeng}\
  \bibnamefont {Chen}}, \bibinfo {author} {\bibfnamefont {Craig}\ \bibnamefont
  {Citro}}, \bibinfo {author} {\bibfnamefont {Greg~S.}\ \bibnamefont
  {Corrado}}, \bibinfo {author} {\bibfnamefont {Andy}\ \bibnamefont {Davis}},
  \bibinfo {author} {\bibfnamefont {Jeffrey}\ \bibnamefont {Dean}}, \bibinfo
  {author} {\bibfnamefont {Matthieu}\ \bibnamefont {Devin}}, \bibinfo {author}
  {\bibfnamefont {Sanjay}\ \bibnamefont {Ghemawat}}, \bibinfo {author}
  {\bibfnamefont {Ian}\ \bibnamefont {Goodfellow}}, \bibinfo {author}
  {\bibfnamefont {Andrew}\ \bibnamefont {Harp}}, \bibinfo {author}
  {\bibfnamefont {Geoffrey}\ \bibnamefont {Irving}}, \bibinfo {author}
  {\bibfnamefont {Michael}\ \bibnamefont {Isard}}, \bibinfo {author}
  {\bibfnamefont {Yangqing}\ \bibnamefont {Jia}}, \bibinfo {author}
  {\bibfnamefont {Rafal}\ \bibnamefont {Jozefowicz}}, \bibinfo {author}
  {\bibfnamefont {Lukasz}\ \bibnamefont {Kaiser}}, \bibinfo {author}
  {\bibfnamefont {Manjunath}\ \bibnamefont {Kudlur}}, \bibinfo {author}
  {\bibfnamefont {Josh}\ \bibnamefont {Levenberg}}, \bibinfo {author}
  {\bibfnamefont {Dandelion}\ \bibnamefont {Man\'{e}}}, \bibinfo {author}
  {\bibfnamefont {Rajat}\ \bibnamefont {Monga}}, \bibinfo {author}
  {\bibfnamefont {Sherry}\ \bibnamefont {Moore}}, \bibinfo {author}
  {\bibfnamefont {Derek}\ \bibnamefont {Murray}}, \bibinfo {author}
  {\bibfnamefont {Chris}\ \bibnamefont {Olah}}, \bibinfo {author}
  {\bibfnamefont {Mike}\ \bibnamefont {Schuster}}, \bibinfo {author}
  {\bibfnamefont {Jonathon}\ \bibnamefont {Shlens}}, \bibinfo {author}
  {\bibfnamefont {Benoit}\ \bibnamefont {Steiner}}, \bibinfo {author}
  {\bibfnamefont {Ilya}\ \bibnamefont {Sutskever}}, \bibinfo {author}
  {\bibfnamefont {Kunal}\ \bibnamefont {Talwar}}, \bibinfo {author}
  {\bibfnamefont {Paul}\ \bibnamefont {Tucker}}, \bibinfo {author}
  {\bibfnamefont {Vincent}\ \bibnamefont {Vanhoucke}}, \bibinfo {author}
  {\bibfnamefont {Vijay}\ \bibnamefont {Vasudevan}}, \bibinfo {author}
  {\bibfnamefont {Fernanda}\ \bibnamefont {Vi\'{e}gas}}, \bibinfo {author}
  {\bibfnamefont {Oriol}\ \bibnamefont {Vinyals}}, \bibinfo {author}
  {\bibfnamefont {Pete}\ \bibnamefont {Warden}}, \bibinfo {author}
  {\bibfnamefont {Martin}\ \bibnamefont {Wattenberg}}, \bibinfo {author}
  {\bibfnamefont {Martin}\ \bibnamefont {Wicke}}, \bibinfo {author}
  {\bibfnamefont {Yuan}\ \bibnamefont {Yu}}, \ and\ \bibinfo {author}
  {\bibfnamefont {Xiaoqiang}\ \bibnamefont {Zheng}},\ }\href
  {https://www.tensorflow.org/} {\enquote {\bibinfo {title} {{TensorFlow}:
  Large-scale machine learning on heterogeneous systems},}\ } (\bibinfo {year}
  {2015}),\ \bibinfo {note} {software available from
  tensorflow.org}\BibitemShut {NoStop}%
\bibitem [{\citenamefont {Kingma}\ and\ \citenamefont
  {Ba}(2014)}]{kingma2014adam}%
  \BibitemOpen
  \bibfield  {author} {\bibinfo {author} {\bibfnamefont {Diederik~P}\
  \bibnamefont {Kingma}}\ and\ \bibinfo {author} {\bibfnamefont {Jimmy}\
  \bibnamefont {Ba}},\ }\bibfield  {title} {\enquote {\bibinfo {title} {Adam: A
  method for stochastic optimization},}\ }\href@noop {} {\bibfield  {journal}
  {\bibinfo  {journal} {arXiv preprint arXiv:1412.6980}\ } (\bibinfo {year}
  {2014})}\BibitemShut {NoStop}%
\bibitem [{\citenamefont {Bottou}(2010)}]{SGD-Bottou}%
  \BibitemOpen
  \bibfield  {author} {\bibinfo {author} {\bibfnamefont {L{\'e}on}\
  \bibnamefont {Bottou}},\ }\bibfield  {title} {\enquote {\bibinfo {title}
  {Large-scale machine learning with stochastic gradient descent},}\ }in\
  \href@noop {} {\emph {\bibinfo {booktitle} {Proceedings of COMPSTAT'2010}}},\
  \bibinfo {editor} {edited by\ \bibinfo {editor} {\bibfnamefont {Yves}\
  \bibnamefont {Lechevallier}}\ and\ \bibinfo {editor} {\bibfnamefont
  {Gilbert}\ \bibnamefont {Saporta}}}\ (\bibinfo  {publisher} {Physica-Verlag
  HD},\ \bibinfo {address} {Heidelberg},\ \bibinfo {year} {2010})\ pp.\
  \bibinfo {pages} {177--186}\BibitemShut {NoStop}%
\bibitem [{\citenamefont {Xu}\ \emph {et~al.}(2015)\citenamefont {Xu},
  \citenamefont {Wang}, \citenamefont {Chen},\ and\ \citenamefont
  {Li}}]{xu2015empirical}%
  \BibitemOpen
  \bibfield  {author} {\bibinfo {author} {\bibfnamefont {Bing}\ \bibnamefont
  {Xu}}, \bibinfo {author} {\bibfnamefont {Naiyan}\ \bibnamefont {Wang}},
  \bibinfo {author} {\bibfnamefont {Tianqi}\ \bibnamefont {Chen}}, \ and\
  \bibinfo {author} {\bibfnamefont {Mu}~\bibnamefont {Li}},\ }\bibfield
  {title} {\enquote {\bibinfo {title} {Empirical evaluation of rectified
  activations in convolutional network},}\ }\href@noop {} {\bibfield  {journal}
  {\bibinfo  {journal} {arXiv preprint arXiv:1505.00853}\ } (\bibinfo {year}
  {2015})}\BibitemShut {NoStop}%
\bibitem [{\citenamefont {Wales}\ and\ \citenamefont
  {Doye}(1997)}]{basin-hopping}%
  \BibitemOpen
  \bibfield  {author} {\bibinfo {author} {\bibfnamefont {David~J.}\
  \bibnamefont {Wales}}\ and\ \bibinfo {author} {\bibfnamefont {Jonathan
  P.~K.}\ \bibnamefont {Doye}},\ }\bibfield  {title} {\enquote {\bibinfo
  {title} {Global optimization by basin-hopping and the lowest energy
  structures of lennard-jones clusters containing up to 110 atoms},}\
  }\bibfield  {booktitle} {\emph {\bibinfo {booktitle} {J. Phys. Chem. A}},\
  }\href {\doibase 10.1021/jp970984n} {\bibfield  {journal} {\bibinfo
  {journal} {J. Phys. Chem. A}\ }\textbf {\bibinfo {volume} {101}},\ \bibinfo
  {pages} {5111--5116} (\bibinfo {year} {1997})}\BibitemShut {NoStop}%
\bibitem [{Note5()}]{Note5}%
  \BibitemOpen
  \bibinfo {note} {Https://www.scipy.org}\BibitemShut {NoStop}%
\bibitem [{\citenamefont {Park}\ \emph {et~al.}(2020)\citenamefont {Park},
  \citenamefont {Park}, \citenamefont {Lee},\ and\ \citenamefont
  {Kim}}]{HPark2020}%
  \BibitemOpen
  \bibfield  {author} {\bibinfo {author} {\bibfnamefont {Hyeong~Seon}\
  \bibnamefont {Park}}, \bibinfo {author} {\bibfnamefont {Seong-Heum}\
  \bibnamefont {Park}}, \bibinfo {author} {\bibfnamefont {Hyunbok}\
  \bibnamefont {Lee}}, \ and\ \bibinfo {author} {\bibfnamefont {Heung-Sik}\
  \bibnamefont {Kim}},\ }\bibfield  {title} {\enquote {\bibinfo {title} {Deep
  learning applied to peak fitting of spectroscopic data in frequency
  domain},}\ }\href {\doibase 10.3938/NPSM.70.920} {\bibfield  {journal}
  {\bibinfo  {journal} {New Phys.: Sae Mulli}\ }\textbf {\bibinfo {volume}
  {70}},\ \bibinfo {pages} {920} (\bibinfo {year} {2020})}\BibitemShut
  {NoStop}%
\end{thebibliography}%


\end{document}